# Geomagnetic origin of the radio emission from cosmic ray induced air showers observed by CODALEMA

D. Ardouin[a], A. Belletoile[a,c], C. Berat[c], D. Breton[d], D. Charrier[a], J. Chauvin[c], M. Chendeb[e], A. Cordier[d], S. Dagoret-Campagne[d], R. Dallier[a], L. Denis[b], C. Dumez-Viou[b], C. Fabrice[b], T. Garçon[a], X. Garrido[d], N. Gautherot[f], T. Gousset[a], F. Haddad[a], D.H. Koang[c], J. Lamblin[a], P. Lautridou[a], D. Lebrun[c], A. Lecacheux[b], F. Lefeuvre[g], L. Martin[a], E. Meyer[f], F. Meyer[f], N. Meyer-Vernet[b], D. Monnier-Ragaigne[d], F. Montanet[c], K. Payet[c], G. Plantier[e], O. Ravel[a], B. Revenu[a], C. Riviere[c], T. Saugrin[a], A. Sourice[e], P. Stassi[c], A. Stutz[c], S. Valcares[a]

**The CODALEMA Collaboration**
[a]SUBATECH, Université de Nantes/Ecole des Mines de Nantes/IN2P3-CNRS, Nantes France.
[b]LESIA, USN de Nançay, Observatoire de Paris-Meudon/INSU-CNRS, Meudon France.
[c]LPSC, Université Joseph Fourier/INPG/IN2P3-CNRS, Grenoble France.
[d]LAL, Université Paris-Sud/IN2P3-CNRS, Orsay France.
[e]GSII, ESEO, Angers France.
[f]LAOB, Université de Besançon/INSU-CNRS, Besançon France.
[g]LPCE, Université d'Orléans/INSU-CNRS, Orléans France.

Abstract : The new setup of the CODALEMA experiment installed at the Radio Observatory in Nançay, France, is described. It includes broadband active dipole antennas and an extended and upgraded particle detector array. The latter gives access to the air shower energy, allowing us to compute the efficiency of the radio array as a function of energy. We also observe a large asymmetry in counting rates between showers coming from the North and the South in spite of the symmetry of the detector. The observed asymmetry can be interpreted as a signature of the geomagnetic origin of the air shower radio emission. A simple linear dependence of the electric field with respect to $\mathbf{v} \wedge \mathbf{B}$ is used which reproduces the angular dependencies of the number of radio events and their electric polarity.

**PACS:**
95.55.Jz; 95.85.Ry; 96.40.-z

**Keywords**
Radio detection; Extensive Air showers; Ultra-High-Energy Cosmic Rays

## 1. Introduction

Cosmic ray induced extensive air showers (EAS) generate radio electric fields that become measurable beyond $10^{16}$eV. Recent observations of this phenomenon were made by CODALEMA [1] and LOPES [2]. Preliminary results were also obtained in Argentina [3,4]. These developments aim at gauging the interest of the radio detection technique of air showers for the field of ultra high energy cosmic ray research. An important milestone resides in the understanding of the signal and its main characteristics, notably the electric field generation mechanism.

Concerning the latter topic, an electron excess in the shower core was first put forward as the source of radio emission by Askar'yan [5]. Later Kahn and Lerche argued that the geomagnetic separation of electrons and positrons provides a more efficient source [6]. This was supported by Allan using a different framework [7]. More recently, several comprehensive frameworks have been proposed [8,9,10]. The most advanced approach by Huege and collaborators [8] puts an emphasis on the geosynchrotron emission and gives a wealth of details on the associated radio emission pattern. Lasty, Scholten and collaborators [10] recently proposed a different approach where the electric transverse current resulting from the electron and positron drift in the geomagnetic field is the driving phenomenon for the radio emission.

The observation of an anisotropy in the number of radio events is presented in this paper. It is shown that it can be reproduced with an emission mechanism where the amplitude of the electric field is proportional to $\mathbf{v} \wedge \mathbf{B}$ , $\mathbf{v}$ being the direction of the shower axis and $\mathbf{B}$ the Earth magnetic field at the location of the experi-ment. Earlier claims of such an observation were drawn out from observed counting rates [11,12] though with a much lower confidence level. All these observations may have a large impact on a future radio array design since a pure polarization proportional to **v**∧**B** rules out the significance of an electric field measurement along the **B** direction.

Section 2 describes the new experimental setup and presents the reconstruction methods. The detection efficiency of the antenna array as a function of the energy is studied in section 3a. Section 3b demonstrates and quantifies the counting rate asymmetry between air showers coming from the North and the South. An interpretation of this observation by a proportionality of the electric field strength to **v**∧**B** is proposed and discussed in section 3c. The observed angular distribution of the electric field polarity is presented in section 3.d. Conclusions and some prospects are given in the last section.

## 2. The experimental setup
### a. The antennas

In the early stage of the CODALEMA experiment, the use of some of the 144 conic logarithmic antennas from the Nançay Decametric Array [13] demonstrated the possible detection of radio signals in coincidence with ground detectors [14, 1]. Fully operational since the 1980's, these antennas are tilted toward the ecliptic plane (20° South in the meridian plane) and are consequently characterized by a slightly asymmetric detection lobe, thus favouring the detection efficiency toward the South [13]. In our first analysis [15] of the main features of the detection method, this was not identified as an annoyance factor though some biased interpretations could have been revealed in a more detailed analysis phase. In other respects, the huge size (6 m high and 5 m wide) of these antennas prevented the development of such units over a larger area.

In the new CODALEMA setup, simplicity, size, cost and performance were the major criteria for the design of a new broadband antenna based on a fat active dipole concept [16]. This dipole antenna is made of two 0.6 m long, 0.1 m wide aluminum slats of 1 mm thickness, separated by a 10 mm gap and is held horizontally above the ground by a 1m high plastic mast. The wires are loaded by a dedicated, high input impedance, low noise (1 nV/√Hz), 36 dB amplifier with a 100 kHz – 220 MHz bandwidth at 3 dB [17]. To avoid possible non-linearity effects due to a 2 GW broadcast local transmitter at 162 kHz, the input of the preamplifier is high pass filtered (20 dB at 162 kHz) resulting in a 1-220 MHz output signal bandwidth at 3 dB. The effective length of the free space antenna is almost constant for low frequencies whereas the directivity gain stays almost isotropic. The antenna radiator length results in a resonating behaviour around 115 MHz. Compared to a wire dipole, a radiator with a small length/thickness ratio has a bigger capacitance (9pF) and a smaller inductance and, consequently, a smaller Q-factor. The antenna resonance is decreased and the antenna losses are thus minimized. Above the resonance, the inductive behavior dominates and the gain decreases. The effective length of the free space antenna is almost constant for frequencies below 25MHz (short dipole) but increases to 7dB in the 10-100MHz band. This variation grows to 19 dB if the antenna is held 1m above a perfect ground plane. In this band, the antenna directivity stays almost isotropic. Validation of this dipole concept was obtained by observing the radio source Cassiopeia A in correlation [17] with the Nançay Decametric Array. Its sensitivity to the galactic noise variations has been deduced from a measurement of the sky background spectrum [18].

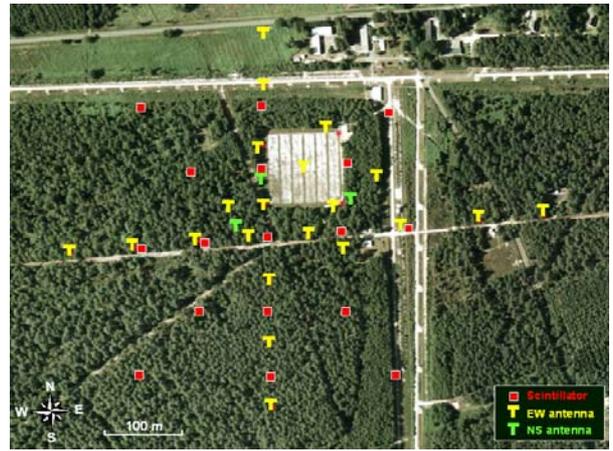

*Figure 1 : Schematic view of the Codalema experimental setup in January 2008 superimposed on an aerial view of the Nançay observatory. Plastic scintillators are depicted as squares. "T" represent dipole antenna oriented in the EW and NS directions. Only EW antennas along the EW and NS main axis of the array are used in this analysis. The large light gray square is the Nançay Decametric Array.*

### b. The two arrays of detectors

Two dedicated overlapping arrays of detectors (figure 1) have been deployed in order to measure simultaneously the particles reaching the ground and the radio signals. Currently, the radio array consists of 24 antennas spaced 90 m from each other, forming a cross with arms of 600 m length. The 14 antennas on the NS and EW arms have been used for the present work. The particle detector array is made of 17 scintillators located on a grid with an approximate spacing of 85 m. It covers a $340 \times 340$ m$^2$ area where the center roughly corresponds to the radio array center [19].

Each particle detector station includes a thick plastic scintillator viewed by two photo-multiplier tubes inserted in a stainless steel box. Each particle detector is weather sheltered by a 1 m$^3$ plastic container. The two photo-multipliers have their high voltage supply set to work at two different gains in order to handle an overall dynamic range from 0.3 to 3000 Vertical Equivalent Muons (VEM).

All the detectors and antennas are wired to a central shelter housing the power supplies, the racks of electronics and the computers for data acquisition. In the standard acquisition mode, the particle detection system acts as a master EAS trigger while the antennas are configured in a slave mode.

Signals from both arrays of detectors are directed to 4 channels 6U VME waveform digitizer boards [20]. The Matacq board performs a fast 12-bit waveform digitalization with a 300 MHz analog bandwidth at a sampling rate set to 1GS/s and in a memory depth of 2560 points (2.5 µs of signal). The maximum range of 1 V on the ADC analogical inputs defines a lower significant bit at 250 µV. The noise of the antenna chain (antenna + preamplifier + cable) measured at the input of the digitizer is less than 200 µV rms.

All the ADC boards are externally triggered by a dedicated 16-fold multiplicity circuit. This circuit discriminates the photomultiplier signals with a threshold corresponding to 0.3 VEM and compares the resulting multiplicity to a remote controlled level. In standard data acquisition conditions, a firing of the 5 central stations within a 600 ns time window is required. This trigger condition leads to an event rate of about 8 events per hour.

### c. Offline data processing

At the first stage of the offline analysis, as explained in detail in previous articles [1, 15], the antenna signals are numerically filtered (23 – 83 MHz) and corrected for the cable frequency response. The relative gains are adjusted using the galactic background. Transient radio pulses indicating the presence of a cosmic ray shower are searched independently in each antenna signal, without the help of a beam forming technique. In the previous CODALEMA phase [1,15], a leading edge discrimination in a bandwidth free of emitters was used. Such a method has the advantage of being simple and very efficient provided the chosen bandwidth remains free from occasional emitters. When dealing with opposite situations, the linear prediction method [21] is helpful. For the current analysis, we have implemented this approach for pulse detection. An estimate $y_e$ for a point n in a time series computed, from a linear combination of the M preceding values

$$y_e(n) = \Sigma^M_{k=1} \ a_k \ y(n-k)$$

has been found appropriate for predicting a signal that shows some regular behaviour. The occurrence of a transient signal in the time series at point $n_0$ is distinguished by a large discrepancy (at least 13 standard deviation) between the recorded value and the estimated one in the vicinity of $n_0$ (typically 2 µs). The pulse time is the bin $n_0$ one. The $a_k$ coefficients are calculated for each event to minimize the deviation

$$\Sigma_{k \in T} \ (y(k) - y_e(k))^2$$

in a training window T. Since it is not a priori known when the pulse occurs in the 2.5 µs time window series, T spans the whole time series. The number of coefficients M depends on the typical radio frequency interference content at the observation location. It was empirically adjusted to M = 250.

Once this tagging procedure is applied and pulses are detected, we can associate each pulse to an absolute time which has been corrected from cable and electronics delays previously measured with a noise generator. From the time information of all antennas, the arrival direction of the shower plane is extracted by triangulation. One should note that a minimum of 3 unaligned tagged antennas is required for defining any arrival direction. With the linear prediction method, the number of reconstructed cosmic ray events (inside the selection criteria described later) increases by 30 % as compared to our previous threshold discrimination technique.

The information on charge and timing of the scintillator detectors are extracted from the digitalized signal by fitting the recorded pulse shape. EAS arrival direction is obtained using the time of flight between different stations, after a signal transit time correction. A plane shower front is assumed.

The number of particles reaching the ground (shower size) and the core position are calculated from the measured particle densities in the detectors. For this process, their positions are projected in the plane perpendicular to the arrival direction. This lateral distribution is fitted with an analytical Nishimura-Kamata-Greisen lateral distribution [22] using a minimization algorithm. The core position and the shower size are extracted and then back projected in the detection plane. The events are sorted into internal and external categories. Internal events are identified by requiring a larger particle density in the internal detectors as compared to the external detectors lying on the sides of the array. This corresponds physically to a shower core contained inside of the array. Simulations have shown that a satisfactory accuracy for both core and size values is thus obtained mainly for internal events while the external ones result in less reliable values. The energy is obtained using the Constant Intensity Cut (CIC) method [23] to define a vertical equivalent shower size $N_0 = N(\theta=0, E)$ with an experimental attenuation length close to 190 g/cm$^2$. From simulations of proton induced EAS run with AIRES [24], the energy is related to $N_0$ by the following formula $E = 2.14 \ 10^{10} \ N_0^{0.9}$, with a resolution of the order of $\Delta E/E \sim 30$ % at $E = 10^{17}$ eV assuming protons as primary particles. The estimation of the shower energy is simulation-dependent and the uncertainty mainly originates from the shower fluctuations and the primary nature.

At the end of the raw data processing, the radio and particle event information are folded into one single event by using the time tagging criteria between the antenna array and the particle detector array (the time dating of the radio and particle events being given by the same time reference service of the Observatory).

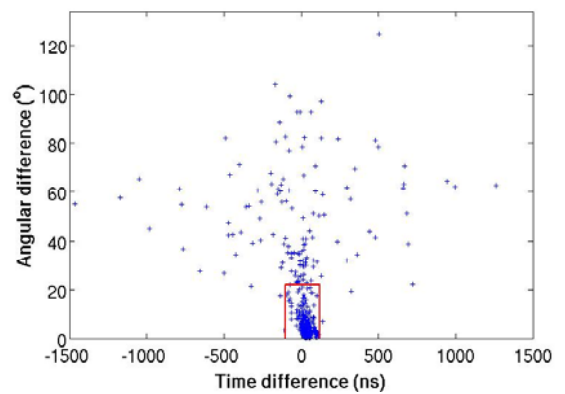

*Figure 2 : Angular difference versus time difference between the reconstructed events from the particle and the radio arrays. The box corresponds to the coincidence criteria.*

## 3. Results

The results hereafter presented have been deduced from a sample of events recorded between November 27th 2006 and March 20th 2008. They correspond to 355 effective days of stable data acquisition. For consistency, information from the antennas recently added has been ignored in this analysis.

### a. Radio detection efficiency

Table 1 summarizes the data accumulated during the above period and gives the number of events seen by the particle array i.e. the number of triggers (scintillators), the number of events detected in coincidence by the antenna array (antennas) and the number of events with a clear correlation with respect to the reconstructed arrival directions (Coincidences). For the latter category, the following criteria must be met: a time coincidence (within ±100 ns) and an angular difference smaller than 20° in the arrival directions as reconstructed from both the particle and radio detectors respectively (figure 2). These numbers are given successively for all types of events, internal events as defined in section 2.c and internal events with estimated energies above $10^{16}$ eV and $10^{16.7}$ eV.

| Event types | Scintillators | Antennas | Coincidences |
|---|---|---|---|
| Reconstructed events | 61517 | 750 | 619 |
| Internal events | 28128 | 195 | 157 |
| Log E >16 | 7889 | 169 | 154 |
| Log E>16.7 | 692 | 134 | 129 |

*Table 1 : Number of events recorded in each array and in coincidence in both arrays. Statistics are given for internal events (see definition in the text) and two energy cuts.*

Figure 2 illustrates the correlation between the event reconstructions from both arrays. There are two improvements as compared to the previous results of CODALEMA [15]. First, mainly due to a recording time of 2.5 μs instead of 10 μs in the previous setup, the number of random events decreases and helps relax the selection criteria. However, we limit the angular difference to 20° in order to keep only well reconstructed events for the analysis reported in the paper.

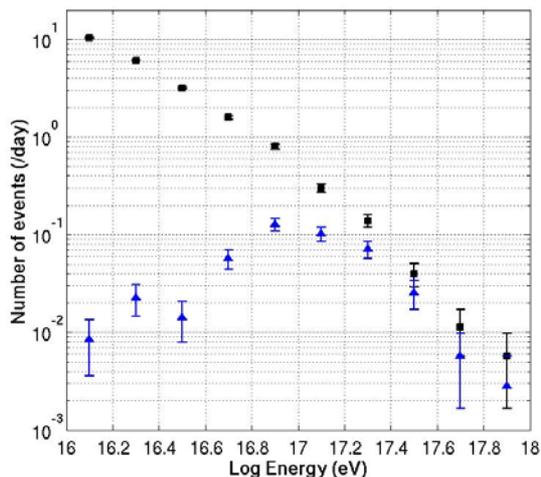

*Figure 3 : Energy distribution of "internal" events measured by the ground particle detector array (squares) and seen in coincidence with the antenna array (triangles).*

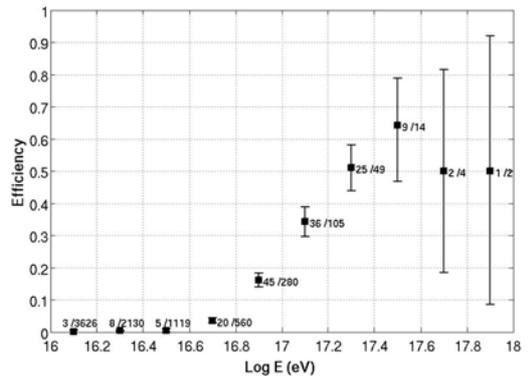

*Figure 4 : Efficiency of the radio detection versus energy deduced from the scintillator analysis.*

About 25% of the events (217 among 891 events) are outside the coincidence window instead of 90% in the previous analysis. We estimate that only few random events are within the coincidence window. Second, with the upgraded particle detector array, 68% of the events have an angular difference below 3.5° in the coincidence window, instead of 6° in the previous analysis. Compared to the previous analysis, the quality of the event reconstruction is improved.

As seen in Table 1, the number of detected events by the particle and radio apparatus differ by almost two orders of magnitude. This is due to a different energy threshold. The energy distribution measured by the ground particle array for internal events is displayed in figure 3 and compared with the same energy distribution for events measured in coincidence by the antenna array. While 1.8 events per day are recorded with the antenna array, the internal events selection decreases this rate to 0.5 event per day.

The threshold behaviour of the radio detector is clearly visible below $10^{17}$ eV. This behaviour has non-trivial consequences for the observations described later. The energy threshold of the ground particle array is far below the range shown in figure 3, around $10^{15}$ eV. Both distributions converge at the highest energies. This reflects the increase of the radio-detection efficiency (figure 4), defined as the ratio of the number of radio detected events over the total number of events. It regularly increases above $3\times10^{16}$ eV and reaches roughly 50 % at $2\times10^{17}$ eV. This efficiency will be discussed again in the section 3.c.

### b. Azimuthal asymmetry

Figure 5 represents the arrival directions of the radio events (known as sky maps) in local coordinates (zenith, azimuth). A striking feature is the shape of the azimuthal distribution, and more specifically the large asymmetry in the observed counting rate between the North (top) and South (bottom) sectors.

The lack of events coming from the southern part of the (local) hemisphere is expected only if the showers coming from North or South generate different radio signals. Note that this South side deficit is not observed on an antenna background (i.e. when the antennas are running in a self trigger mode).

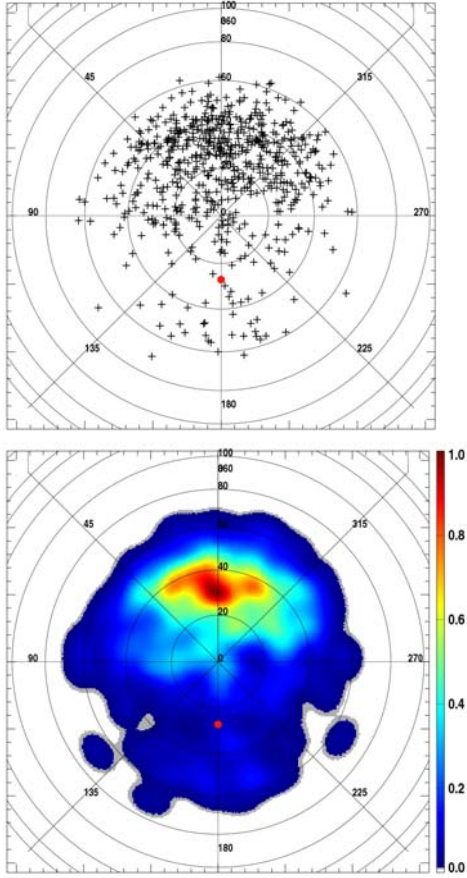

*Figure 5 : Sky maps of observed radio events. Raw event sky map (top) and 10° gaussian smoothed map (bottom) are shown. The zenith is at the center, the azimuth is: North (top, 0°), West (left, 90°), South (bottom, 180°) and East (right, 270°); the direction of the geomagnetic field at Nançay is indicated by the dot.*

In order to characterize the North-South asymmetry, we consider the ratio $n_{South}/n_{Tot}$ of the number of events coming from the South (90° < φ < 270°) to the total number of events without additional selection. The observed ratio is 109/619 = 0.17 ± 0.02. In order to quantify the asymmetry level, we can compare this ratio to the value obtained from simulated events drawn from a symmetrized version of the coverage map. This map is constructed from a fit of the whole data set zenith distribution and a fit of the azimuthal distribution of the northern region mirrored into the southern region. With this coverage, we have a ratio of 0.5 which is expected by construction. The observed deviation from symmetry is thus 0.5-0.17=0.33 which corresponds to 0.33/0.02=16 standard deviations. Moreover, the experimental value $n_{South}/n_{Tot}$ is stable in time as shown in the figure 6 where it has been computed for 7 independent time ordered samples of events (6 with 88 events and 1 with 91 events). This observation cannot therefore be associated to a detector failure or a statistical fluctuation. The corresponding ratio $n_{East}/n_{Tot}$ estimated for the eastern events (180° < φ < 360°) shows values compatible with symmetric simulations (see figure 6).

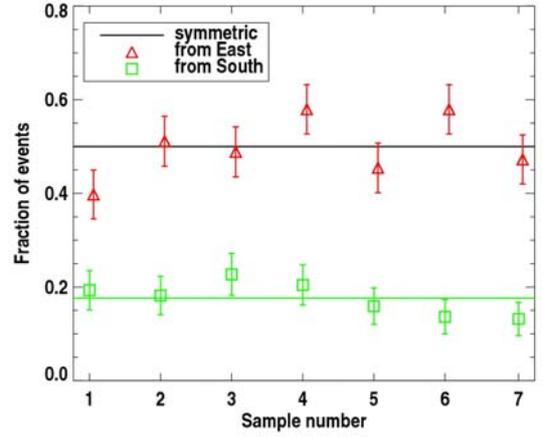

*Figure 6 : Fraction of events for 7 independent samples of events (619 events in total). The fractions of events coming from the East and from the South are indicated by triangles and squares respectively. The expected ratio of 0.5 in the symmetric case is indicated.*

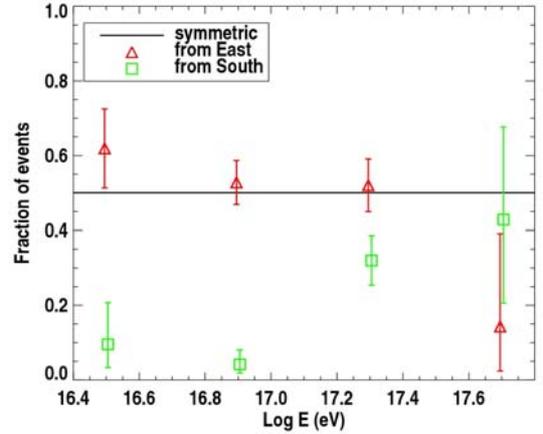

*Figure 7 : Evolution of the fraction of events (squares: coming from the East, triangles: coming from the South) with energy. The expected ratio of 0.5 in the symmetric case is indicated.*

Now that the global asymmetry has been established and quantified, we can investigate its dependence with the energy using the internal events. Figure 7 shows that the value of the ratio $n_{South}/n_{Tot}$ becomes compatible with 0.5 (as expected for a symmetric distribution) when energy increases, showing that the asymmetry in counting rates is a threshold effect. At low energy, the electric fields are close to the detection threshold and a difference in magnitude implies a difference in counting rate. At high energy, the overall strengthening of the radio pulses washes out the latter difference and any asymmetry has to be searched for in the signal amplitude. Observation of such an effect is a more challenging task on which we are currently working. There is no variation of the ratio $n_{East}/n_{Tot}$ as a function of energy (see figure 7): eastern events account for 50% of the flux at close to and well beyond threshold, showing that no asymmetry exists between West and East fields generated by the air showers.

The zenithal and azimuthal distributions observed for the internal scintillator events above $10^{17}$ eV are presented in figure 8 (top and bottom). The zenithal distribution is fitted by the following empirical function:

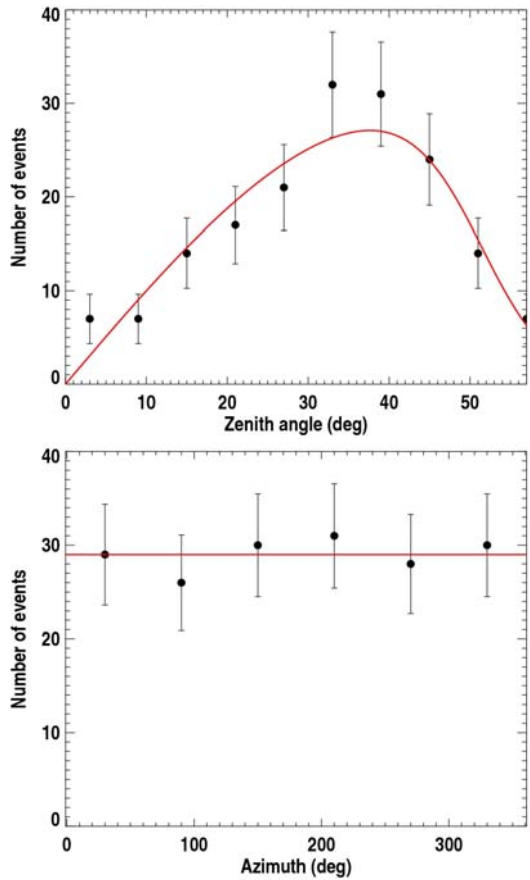

*Figure 8 : Zenith (top) and azimuthal (bottom) angular distributions of internal showers with $E > 10^{17} eV$ detected by the particle detector array used as a trigger.*

$$dN/d\theta=(a+b\theta)\cos(\theta)\sin(\theta)/(1+\exp((\theta-\theta_0)/\theta_1))$$

.

The Fermi-Dirac function accounts for the atmospheric attenuation and the scintillator thinness. The $\cos(\theta)\sin(\theta)$ term describes the cosmic flux. The linear term is a second order adjustment of the order of few percents. The free parameters are found to be a=44.96, b=0.57, $\theta_0$=49.18 degrees and $\theta_1$=5.14 degrees. The azimuthal distribution is compatible with a uniform distribution (figure 8 bottom). Because the particle trigger introduces no significant bias in the azimuthal distribution, any feature observed in the azimuth for radio events should be attributed to the radio signals themselves. An East and West attenuation is expected by construction due to the East/West orientation of the dipole axis – the gain being smaller in these directions – and this property breaks the azimuthal symmetry. However, both North-South and East-West symmetries are preserved with the dipolar antenna used.

One should note that such an azimuthal asymmetry has not been observed with the previous CODALEMA setup [15]. As quoted above, log-periodic (conical) antennas with axis tilted by 20° to the South [14] were used in that setup. The directivity was maximum in the cone axis direction and the radio detection of events coming from the North was disfavored. The overall detector was not North-South symmetric and as a consequence it was probably not suited for a direct measurement of any North-South asymmetry of physical origin. Unfortunately, due to a lack of statistics, a closer inspection of the data of Ref.[15], especially searching for differences in the azimuthal direction for bins with comparable gain (e.g., comparing S, SE and SW sectors) does not allow us to confirm or disprove the observation made in the present study. It should be noted that that kind of spiral antenna has been designed to measure circular polarization (left or right) of the electric field that spreads almost vertically toward them [14]. The instrumental response of these sensors to a linearly polarized field could possibly mask some azimuthal anisotropies, which are precisely associated to linearly polarized radio signals.

### c. Physical interpretation

The observed North-South asymmetry is clear and unambiguous. It calls for an investigation of the electric field generation mechanism when air showers develop in the atmosphere. With the observed pattern, an obvious candidate for symmetry breaking effect in the electric field generation is the geomagnetic field. Since the Lorentz force acting on the charged particles is at the origin of different emission mechanisms of the electric field, involving the geomagnetic field (geo-synchrotron and macroscopic approaches), the electric field magnitude itself should depend on the values of the vector cross product $\mathbf{v} \wedge \mathbf{B}$. In addition to the electric field magnitude, the output signal also depends on the polarization. We will then make the assumption that this polarization is linear and oriented along $\mathbf{v} \wedge \mathbf{B}$. Not all geomagnetic induced fields have this property, but this is true in the geosynchrotron approach at small impact parameters [25] and in the transverse current approach [10]. Overall we will consider that the signal amplitude given by the EW dipole is proportional to $|(\mathbf{v} \wedge \mathbf{B})_{EW}|$, which is the magnitude of the vector cross product projected on the East-West axis (the orientation of the antennas).

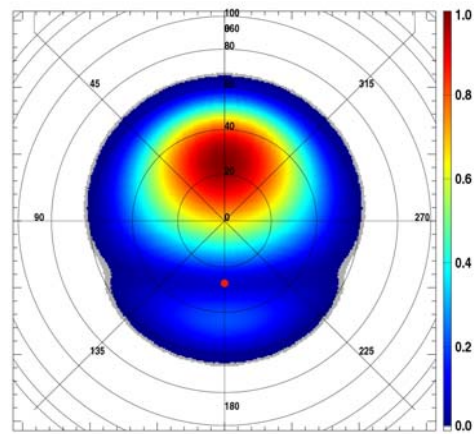

*Figure 9 : Sky map calculated by considering the EW component of the Lorentz force multiplied by the trigger coverage map. The color scale is normalized to 1 in the direction of the maximum.*

In order to compute a density map giving the expected number of radio events (the so-called coverage map), we need an extra assumption about the relation between the signal magnitude and the true detection of an event. Clearly a stronger signal is easier to detect. We test the simple possibility that the radio efficiency is proportional to the signal magnitude as defined above (later discussed on the figure 11). The predicted event sky map (figure 9) can finally be computed multiplying this efficiency by the trigger coverage map. The trigger coverage map is obtained using the parameterization of the zenithal distribution of the ground detector events above $10^{17}$ eV, presented in the previous subsection, associated with a uniform azimuthal distribution.

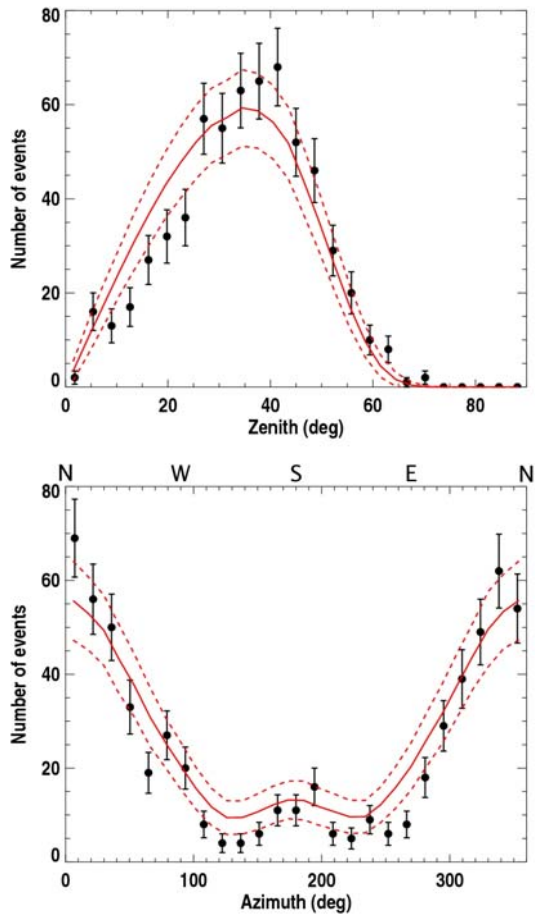

*Figure 10 : Zenith (top) and azimuthal (bottom) angular distributions (black crosses) observed for the radio events. The solid line represents the predicted distribution obtained from simulated events according to the coverage map presented in figure 9. The dashed lines define the +/- 1σ band around the prediction.*

Under these assumptions, figure 9 can be interpreted as a prediction of the shower arrival direction sky map. This prediction appears to be very similar to the observed sky map (figure 5). It reproduces the main features of the experimental distribution: a maximum towards the North with bean-shaped contour lines, a local maximum towards the South and minima in the East and West directions. To check this similarity, a set of events with a size equal to the data set is built by randomly generating arrival directions following the predicted coverage map. Simulated zenith and azimuth angular distributions are then compared to the observed ones in figure 10. The agreement is noticeable for both distributions; especially the relative amplitudes of the two maxima in the azimuthal distribution are fairly well reproduced.

Figure 11 exhibits the efficiency of the radio detection for internal events, i.e. the number of radio detected events divided by the number of triggered events, above $10^{17}$eV plotted as a function of $|(\mathbf{v} \wedge \mathbf{B})_{EW}|/(vB)$. The observed trend confirms the hypothesis that the number of detected events is strongly linked to the electric field. It is also seen that the efficiency seems to vary linearly with the electric field. This can be highlighted only because the present analysis is made around the energy threshold of the experiment. Working at lower or higher energy changes the shape of the plot. This is indirectly shown in figure 12 which now represents the mean energy of radio detected events for each bin of the Lorentz force EW component. To be detected, events with a low value of the Lorentz force EW component, for example coming from East or West, must have a higher energy than events with high values, coming from North. But higher energy events will be detected no matter their arrival direction and the radio efficiency will become independent of the vector cross product.

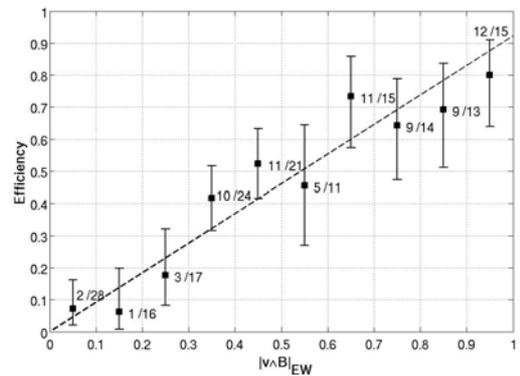

*Figure 11 : Number of radio events relative to the number of scintillator events (E > $10^{17}$ eV) with respect to $|(\mathbf{v} \wedge \mathbf{B})_{EW}|/(vB)$.*

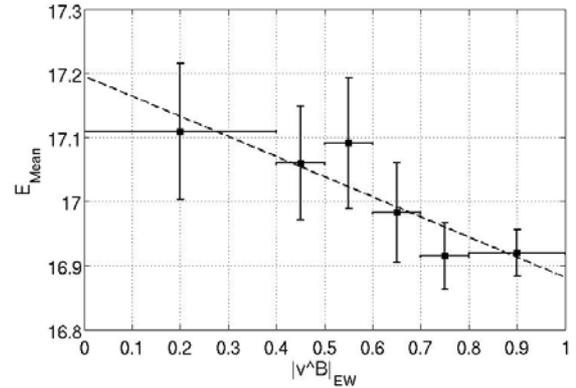

*Figure 12 : Mean energy of showers seen by the radio method versus $|(\mathbf{v} \wedge \mathbf{B})_{EW}|/(vB)$.*

Finally, from these results, it is interesting to return to figure 4. Instead of drawing the radio detection efficiency as a function of the energy, the efficiency is now plotted in figure 13 as a function of the energy weighted by the EW component of the vector cross product E'=Energy$|(\mathbf{v} \wedge \mathbf{B})_{EW}|/(vB)$. The efficiency now increases with increasing E' values up to values close to unity at $10^{17.4}$ eV. The efficiency of 0.5 at $2\times10^{17}$ eV seen in figure 4 is explained by events with low values of the vector cross product EW component, i.e. events coming from East and West, for which the electric field is too weak to be detected by the CODALEMA setup.

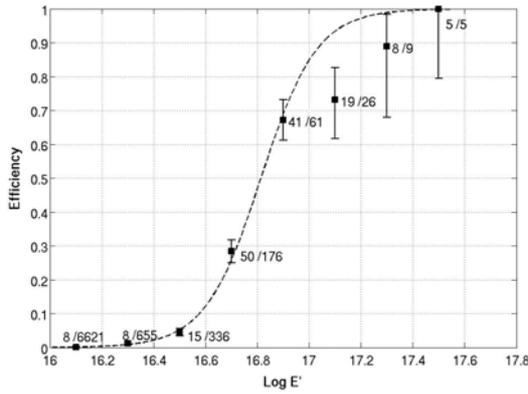

*Figure 13 : Radio detection efficiency versus E' defined as the energy multiplied by $|(v \wedge B)_{EW}|/(vB)$. The dashed line corresponds to a Fermi function fit.*

The $v \wedge B$ dependence incorporates a linear polarization of the electric field in the direction of the cross vector product. It means that the CODALEMA array (using EW oriented dipoles) measures mainly showers arriving in the orthogonal directions (i.e. North/South). Consequently, an array of NS oriented dipoles should be sensitive to showers arriving from East and West directions. Preliminary results of radio detected showers observed with a new sub-array of three NS oriented dipoles seems to confirm this $v \wedge B$ electric field dependence. Indeed, with the limited set of data collected so far with the NS antenna (45 events), one observes mainly showers arriving from eastern and western directions.

### d. Field polarity

The observed $v \wedge B$ polarization dependence also implies a sign dependence of the field components to the arrival direction of the air shower. The normalized EW component of $v \wedge B$ multiplied by the trigger coverage map is represented in the figure 14 (the same as the figure 9 besides the sign). This sky map should be interpreted as the event density for each polarity and can be directly compared to the observed events. Positive signals are arbitrarily associated to showers coming from Northern directions while events in the South hemisphere are characterised by negative pulses.

The pass-band filter used in the data analysis transforms transient signals associated to EAS into multipolar oscillating signals. The sign of the filtered signal extrema is used as the electric field polarity estimate. In order to limit the effect of the noise which sometimes changes the sign of the extremum, we have defined the sign for one event as the majority sign among all the signals associated with tagged antennas. In addition, only events with a clear sign majority (at least a majority of 2 units) are considered. About 2/3 of the events fulfil this condition. The resulting experimental sky map is shown on the figure 15.

The signals seen in the EW polarisation and coming from North are mainly positive while signals coming from South are mainly negative. The agreement with the field polarity distribution shown in figure 14 approaches 86%. The 14 % of events which do not match with the sign of $v \wedge B$ could be due to an emission not always purely polarized as $v \wedge B$. How-ever, one should stress that most of the latter events correspond to a lower signal to noise ratio and consequently a less confident sign determination.

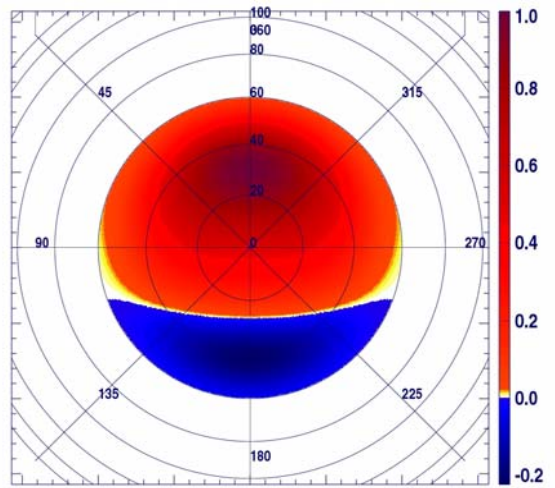

*Figure 14 : Sky map of the predicted polarity of the electric field calculated by considering the EW component of the Lorentz force and multiplied by the trigger coverage map (positive signals are arbitrarily associated to showers coming from Northern directions).*

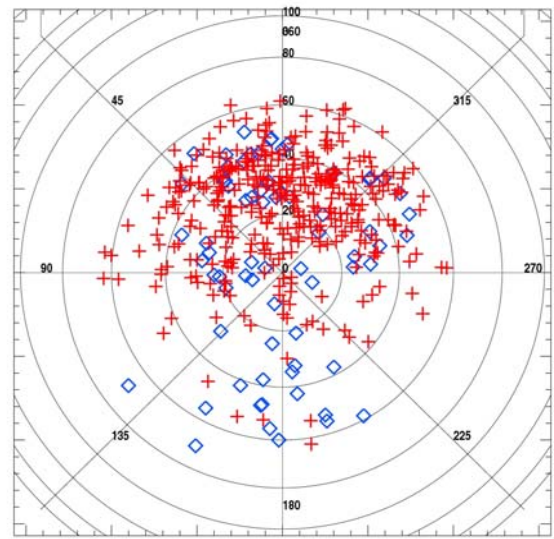

*Figure 15 : Sky map of the signal sign (crosses for positive signs, diamonds for negative signs) of observed radio events.*

The experimental sign of the signal is completely consistent with an emission mechanism with an electric field proportional to $v \wedge B$. Preliminary analysis of the signal polarity deduced from the new NS antenna supports also this conclusion.

### 4. Conclusions and outlook

The first results of the CODALEMA experiment with the new setup (using only active dipoles) confirm the field characteristics extracted from our previous measurement [15]. The addition of a dense array of particle detectors makes the estimation of the shower energy possible and allows CODALEMA to demonstrate the correlation between the radio-detection efficiency and the shower energy. Using the detection of only one EW polarization state, another new result is the apparent depletion of the number of radio detected events from Southern directions. At this stage of our understanding, two conclusions can be drawn: first, the behavior of the measured electric field can be well reproduced by simply considering the vector cross product **v**∧**B** of the Lorentz force; second, the radio signal induced by the geomagnetic field is dominant with the current CODALEMA observation conditions.

These results are consistent with the Allan's conclusion [7] suggesting an electric field oriented along the **v**∧**B** direction and proportional to $|\sin \alpha | = |$**v**∧**B**$|/(vB)$. The LOPES collaboration parameterized the radio signals from the EW polarized antennas in a different manner [26] namely by inserting a fitting factor $(1.16(\pm 0.025) - \cos \alpha)$. In a geomagnetic type approach of the radio emission of EAS, the vector cross product **v**∧**B** here proposed has the advantage of being simple and natural while already including all three NS, EW and vertical components of the electric field. It also explains the sign of the signals and the presence of a secondary maximum for southern events, which does not appear with the LOPES parameterization.

In the near future, a deeper understanding of the polarization effect is mandatory. This should help to distinguish between different microscopic approaches such as geosynchrotron or transverse current models. Because this requires a more comprehensive measurement of all states of polarization, several directions will be investigated within CODALEMA. The new NS oriented dipoles completed by autonomous antenna stations implemented by the end of the year will be analyzed. Additional information will be provided by the use of the Nançay Decametric Array. This apparatus has been recently equipped with the electronics adapted to the transient waveforms characterization. Thanks to the detection of the circular polarization, it will allow to measure at once the full horizontal polarization. The exploitation of these data recorded in coincidence with the particle and dipole arrays should allow a new qualitative step in the interpretation of the radio emission mechanisms.

Consequently, one of the remaining issues is to determine how many and which polarization states have to be measured in order to design an efficient detection method. Indeed, a pure dependence of the radio signal to **v**∧**B** implies proportionality between the NS and the vertical components and thus redundant information. This will impact the cost, the ease of use, and therefore the deployment of the radio technique on a large scale.

**Acknowledgment :**
The authors acknowledge the support from the French Agence Nationale de la Recherche, under the grant ANR-NT05-2_42808.